\documentclass[twocolumn,twocolappendix]{aastex631}
\usepackage{graphicx}
\usepackage{grffile}
\usepackage{dcolumn}
\usepackage{xcolor}
\usepackage{color}
\usepackage{bm}
\usepackage{amssymb, amsmath}

\usepackage{natbib}
\usepackage{soul}
\usepackage{times}

\newcommand{\eg}{e.g.,~}
\newcommand{\ie}{i.e.,~}

\usepackage{cancel}

\begin{document}
	
\preprint{XXX}

%
\title{Electromagnetic Energy Extraction from Kerr Black Holes: Ab-Initio Calculations}

\author[0000-0001-8694-3058]{Claudio Meringolo} \affiliation{Institut f\"ur
  Theoretische Physik, Goethe Universit\"at, Max-von-Laue-Str. 1, D-60438
  Frankfurt am Main, Germany}

\author[0000-0003-0412-0491]{Filippo Camilloni} \affiliation{Institut f\"ur
	Theoretische Physik, Goethe Universit\"at, Max-von-Laue-Str. 1, D-60438
	Frankfurt am Main, Germany}
	
\author[0000-0002-1330-7103]{Luciano Rezzolla} \affiliation{Institut f\"ur
	Theoretische Physik, Goethe Universit\"at, Max-von-Laue-Str. 1, D-60438
	Frankfurt am Main, Germany} \affiliation{CERN, Theoretical Physics
	Department, 1211 Geneva 23, Switzerland} \affiliation{School of
	Mathematics, Trinity College, Dublin, Ireland}




\begin{abstract}
  The possibility of extracting energy from a rotating black hole
  via the Blandford-Znajek mechanism represents a cornerstone of
  relativistic astrophysics. We present general-relativistic
  collisionless kinetic simulations of Kerr black-hole magnetospheres
  covering a wide range in the black-hole spin. Considering a classical
  split-monopole magnetic field, we can reproduce with these ab-initio
  calculations the force-free electrodynamics of rotating black holes and
  measure the power of the jet launched as a function of the spin. The
  Blandford-Znajek luminosity we find is in very good agreement with
  analytic calculations and compatible with general-relativistic
  magnetohydrodynamics simulations via a simple rescaling. These results
  provide strong evidence of the robustness of the Blandford-Znajek
  mechanism and accurate estimates of the electromagnetic luminosity to
  be expected in those scenarios involving rotating black holes across
  the mass scale.
\end{abstract}



\section{Introduction}
\label{intro}

A rotating (or Kerr) black hole (BH) has a reservoir of rotational energy
that can be tapped to drive ultrarelativistic jets if the BH is embedded
in a magnetic field. This mechanism, first proposed by Blandford and
Znajek (BZ)~\citep{Blandford1977} could be at work in accreting BHs across
the mass scale, \ie either in supermassive BHs where it would be
responsible for the phenomenology of Active Galactic Nuclei (AGNs) or in
stellar-mass BHs, where it could represent the engine powering gamma-ray
bursts (GRBs; see, \eg~\cite{Abramowicz2013}). What these apparently
different scenarios have in common is the presence of electromagnetic
(EM) fields building up a magnetosphere and of a force-free plasma
permeating it.

Over the last decades, an enormous literature has been built to describe
and validate the BZ mechanism, either analytically~\citep{Blandford1977,
  Thorne82, Tanabe2008, Morozova2013, Lasota2014, Kinoshita2017,
  Armas2020, Camilloni2022, Toma2025}, via general-relativistic magnetohydrodynamic
(GRMHD) simulations~\citep{Komissarov2001, Komissarov2004b, McKinney04,
  Komissarov2005, Tchekhovskoy2010, Tchekhovskoy2011, Rezzolla:2011,
  Ruiz:2012te, Penna2013, Nathanail2014, Akiyama2019_L1_etal,
  Akiyama2019_L5_etal, Nathanail2016, Gottlieb2021, EHT_SgrA_PaperI_etal,
  EHT_SgrA_PaperV_etal, Shibata2024}, or using more fundamental
particle-based, relativistic simulations~\citep{Parfrey2019, ElMellah2021,
  Hirotani2021, Galishnikova23, Torres2024, Yuan2025, Chen2025}. The two numerical
approaches have distinct advantages, with the fluid nature of the former
ones being particular suited to provide a global and large-scale
description of the launching of magnetically dominated jets and the
kinetic nature of the latter being more effective in reproducing the
physical conditions expected in BH magnetospheres, where the plasma is
closer to a force-free regime. Both approaches, however, have had the
merit of confirming the plausibility of the BZ mechanism under a variety
of conditions.

Exploiting a newly developed general-relativistic particle-in-cell
(GRPIC) numerical code, \texttt{FPIC}, we present the first and most
comprehensive campaign of fully kinetic, two-dimensional simulations of
force-free BH magnetospheres. Using as a reference the classic
split-monopole proposed by Blandford and Znajek~\citep{Blandford1977}, we
simulate the highly dynamic current sheet that forms in the equatorial
plane and that leads to a rich and powerful process of magnetic
reconnection~\citep{Crinquand2022, Ripperda2022, ElMellah2021}. The latter
occurs through the development of the tearing instability, so that a
chain of plasmoids forms along the reconnection layer, which are ejected
at relativistic velocities~\citep{Uzdensky2007,Bhattacharjee2009}.  Our
analysis allows us to characterise the magnetic-reconnection rate and to
express it analytically in terms of the distance from the event horizon
and of the BH spin. More importantly, we quantify the extraction of
rotational energy from the BH to the EM fields and particles via the BZ
mechanism and compare it with the original lowest-order expression by
Blandford and Znajek~\citep{Blandford1977}, with more recent high-order
analytic expressions~\citep{Camilloni2022}, and with GRMHD
simulations~\citep{Tchekhovskoy2010}. In addition to finding a
surprisingly good agreement between the results of the GRPIC simulations
and the high-order analytic estimates, we discuss how the GRMHD results
can also be recovered via a simple rescaling that depends only on the
effective magnetic-field topology. Finally, we show that populations of
particles with negative energy at infinity are present inside the
ergosphere, suggesting that also the Penrose process is
active~\citep{Penrose1971, Dhurandhar:1984, Wagh1985, Asenjo2015, Comisso2021,
  Camilloni2025}.

\section{Results}

\subsection{Overview of the Simulations}

Since the BH mass $M$ plays here only the role of a scaling factor, the
typical scenario characterising a force-free magnetosphere around
rotating BH in general relativity is described by a single parameter,
namely, the dimensionless BH spin $a_*:=J_*/M^2$, where $J_*$ is the spin
angular momentum of the Kerr spacetime. Furthermore, since we are here
interested to perform a systematic study of the dynamics of charged
particles and EM fields in this scenario, we have carried out twelve,
high-resolution, kinetic GRPIC simulations differing in the dimensionless
BH spin (see Table~\ref{tab:table1} for a summary of the
simulations). While they have represented a significant computational
cost, the campaign of simulations has been essential to obtain a clear
and comprehensive description.

The details of the numerical code employed for the simulations are
briefly summarised in the \hyperref[sec:appendix]{Appendix},
but it is important to recall here that all simulations are initiated in
electrovacuum, \ie in the absence of charged particles, and share the
same initial seed magnetic field given by a split-monopole, solution of
the Maxwell equations in the electrovacuum Schwarzschild spacetime. In
this case, using spherical coordinates, the radial component of the
magnetic field as measured by a fiducial observer (FIDO) is given by $B^r
:= \partial_\theta\Psi/\sqrt{\gamma}$, where $\gamma$ is the determinant
of the spatial three-metric $\gamma_{ij}$, while $\Psi(r,\theta)$ is the
gauge-invariant magnetic-flux function $\Psi(r,\theta) = 4M^2B_0
\zeta(\theta)(1-\cos\theta)$. The factor $\zeta(\theta)$ is introduced to
model the equatorial current-sheet necessary to sustain the field
discontinuity~\citep{Gralla2014}. We note that, in ideal conditions, one
would consider a step-function $\zeta=2\mathcal{H}(\theta-\pi/2)-1$,
where $\mathcal{H}(x)$ is the Heaviside function. In practice, we employ
the smooth function $\zeta := 2/\pi\arctan\left[\chi
  (\theta-\pi/2)\right]$ with $\chi$ controlling the jump and fixed at
$\chi = 1000$.

In all simulations, the initial magnetic-field strength $B_0$ is
specified in terms of the dimenionless field $\tilde{B}_0 := r_{g}
/r_{_{\rm L}} = GM\,eB_0 /(m_e c^4)$ where $r_{g}:=GM/c^2$ is the
gravitational radius, $r_{_{\rm L}} := m_e c^2/(eB_0)$ is the Larmor
radius, and $m_e$, $e$ are the reference electron mass and charge,
respectively. Following~\cite{Parfrey2019}, we set $\tilde{B}_0 = 10^3$,
which is equivalent to a magnetic field of $\simeq 1\,{\rm G}$ for a BH
of mass $M\simeq 10\,M_{\odot}$ (or of $\simeq 10^{-5}\,{\rm G}$ for a BH
with the mass of Sgr~A*, \ie $M\simeq 10^{6}\,M_{\odot}$). While these
fields are admittedly not realistic, they keep the computational costs
affordable in our extensive spin exploration and represent a common
choice~\citep{Parfrey2019, Chen2025}.

\begin{figure*}
  \centering
  \includegraphics[width=1\textwidth]{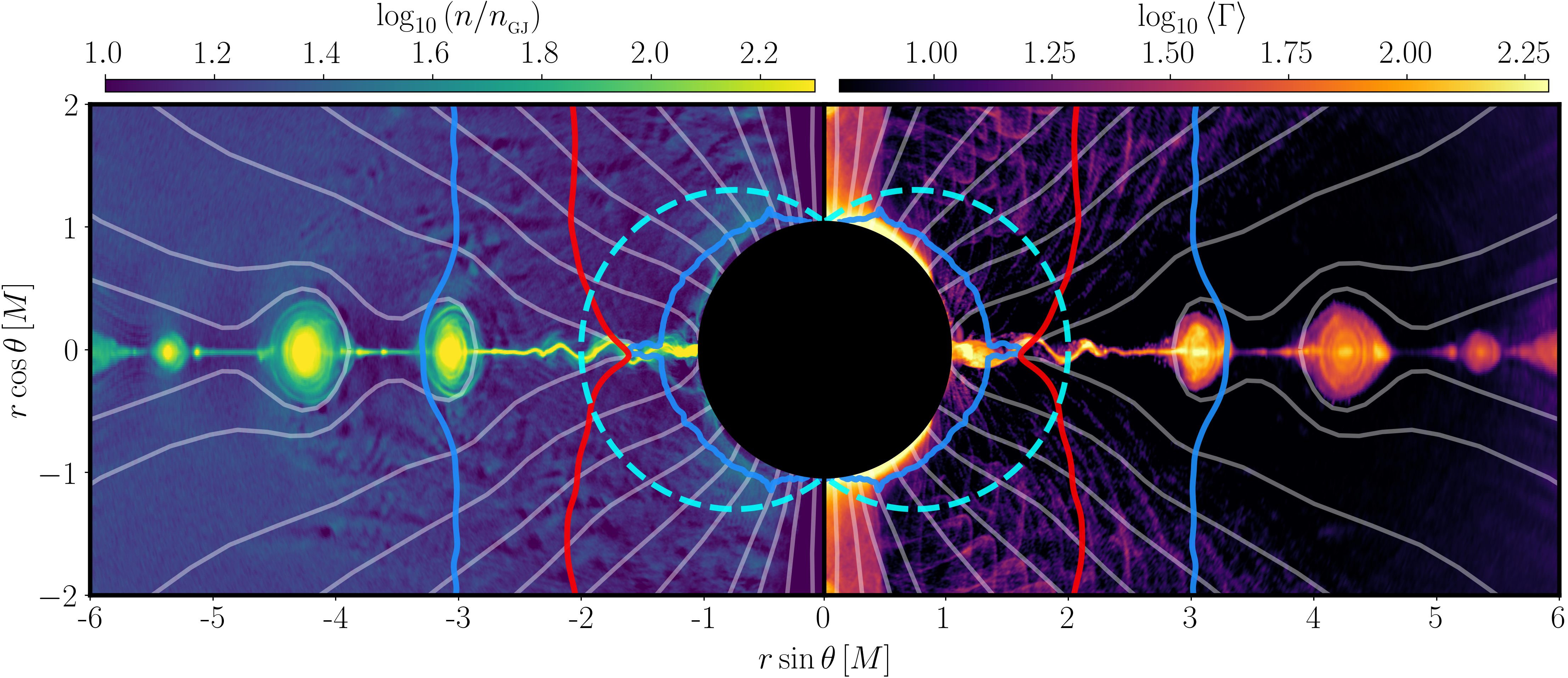}
  \caption{\textit{Left panel:} Total number density of particles
    normalised to the Goldreich-Julian density $n/n_{_{\rm GJ}}$ as
    measured by a FIDO observer. The data refers to time $\bar{t} =
    10\,M$ and to a BH with spin $a_*=0.999$. Also shown are the
    ergosphere (cyan dashed line), the inner and outer light surfaces
    (blue solid lines), and the stagnation surface (red solid
    line). \textit{Right panel:} the same as on the left for but for the
    Lorentz factor $\Gamma$.}
\label{fig:fig1}
\end{figure*}

The value of $B_0$ also provides a reference scale for a number of
plasma-related quantities, such as the Goldreich-Julian number density
$n_{_{\rm GJ}} := \Omega_{h} B_0/(4 \pi c e)$~\citep{Goldreich:1969},
which measures the minimum number density required to screen the
longitudinal components of the electric field, and where $\Omega_{h}$ is
the angular velocity of the event horizon (clearly, higher values of
$\tilde{B}_0$ would force the use of a larger number of particles). Our
setup also preserves the important hierarchy of astrophysical scales,
both in space, \ie $r_{_{\rm L}} \ll d_{p} \ll r_{g}$, and in time, \ie
$\Omega_{h} \ll \omega_{p} \ll \omega_{_{\rm L}}$, where $d_{p} :=
c/ \omega_{p}$ is the plasma skin depth, $\omega_{p} := \sqrt{4 \pi
  n_{_{\rm GJ}}e^2/(m_ec^2)}$ is the plasma frequency, and $\omega_{_{\rm
    L}} := c/r_{_{\rm L}}$ the Larmor frequency. Hereafter, and unless
specified differently, we use geometrised units in which $c=1=G$, with
$c$ and $G$ the speed of light and the gravitational constant,
respectively.

In the presence of spacetime rotation, the initial static Schwarzschild
split-monopole evolves towards a stationary electrovacuum configuration
that preserves a monopolar topology in the poloidal plane and an
equatorial current-sheet, while exhibiting gravitationally-induced
toroidal magnetic fields, as well as electric fields along the
magnetic-field lines. These gravitationally induced EM fields are similar
to what observed in the Wald solution~\citep{Wald:74bh} as considered, for
instance, in \cite{Moesta:2009, Alic:2012, Parfrey2019}.

After the vacuum EM fields settle to a stationary configuration at
$t=t_{\rm inj} = 50\,M$, a $e^-/e^+$-pair plasma is injected in the
spherical volume with coordinate radius $r_{h} < r < 15\,M$, where $r_{h}
:= (1+\sqrt{1-a_*^2})\,M$ marks the location of the event horizon. This
particle injection, that is customary in PIC simulations, is repeated for
all subsequent times $\bar{t} := t - t_{\rm inj}>0$ and it is necessary
to establish a plasma-filled magnetosphere in which the electric-field
components longitudinal to the magnetic-field are screened by the
plasma. Furthermore, in order for the magnetosphere to attain the
force-free regime in which the inertial and thermal contributions of the
particles to the total energy-and-momentum balance are smaller than those
associated with the magnetic field, the condition for the plasma
replenishment is expressed as $n < \mathcal{M}\, n_{_{\rm
    GJ}}$~\citep{Beskin1997}, where $n$ is the total number density $n:=
n_{e^-} + n_{e^+}$, and $n_{e^-}, n_{e^+}$ the number densities of
electrons and positrons, respectively. The factor $\mathcal{M}>1$ is the
so-called multiplicity that we set to $\mathcal{M}=10$ so as to have an
accurate representation of the BH magnetosphere. We have verified that,
in this way, the highly magnetised magnetosphere in our simulations obeys
$\sigma = B_iB^i/(4 \pi n m_e) \gg \Gamma$ ~\citep{Beskin1997}, where
$\sigma$ is the magnetisation parameter, and $\Gamma$ the Lorentz factor
relative to the FIDO; typical values in our simulations are $\sigma
\simeq 100$ and $\sigma/\Gamma \gtrsim 10$.

The particles are injected in each cell with FIDO-frame velocities
randomly drawn from a relativistic Maxwell--J\"uttner of dimensionless
temperature $\Theta := k_\text{B} T/m_e = 0.5$, with $k_\text{B}$ being
the Boltzmann constant. Furthermore, to increase the density of particles
near the equatorial region, we modulate the injected number density with
a simple polar profile $n_{\text{inj}} = n_0 \sin(\theta)$, where the values
of $n_0/n_{_{\rm GJ}}$ depend on the spin and can be found in Table~\ref{tab:table1}. 
We have checked that other injection methods, \eg using the local parallel
electric field as a proxy for the electron-positron
discharge~\citep{Parfrey2019, Bransgrove2021}, lead to results that are
qualitatively and quantitatively similar.

\subsection{Plasmoid formation and dynamics}

Figure~\ref{fig:fig1} offers a representative view at time
$\bar{t}=10\,M$ of the most extreme of our configurations, namely, of a
Kerr BH with spin $a_*=0.999$ (see case \texttt{a.999} in
Table~\ref{tab:table1}). In particular, shown in the left panel of
Fig.~\ref{fig:fig1} is the distribution of the total particle number
density normalised to the Goldreich-Julian value $n_{_{\rm GJ}}$. Note
the presence on the equatorial plane of a current sheet -- where the
poloidal magnetic field inverts its polarity -- and of a series of
plasmoids -- \ie confined plasma concentrations -- moving either to
larger radii or towards the BH horizon. Furthermore, the colormap reveals
that the number density along the current sheet and inside the plasmoids
is on average at least one order of magnitude larger than elsewhere in
the computational domain. Also shown with thin grey lines are the
projections of the magnetic-field lines over the poloidal plane, which
preserve the overall split-monopole topology despite the appearance of
polar and azimuthal components. We recall, in fact, that when the system
reaches force-free quasi-stationary conditions, the magnetic-field lines
rotate rigidly with an angular velocity $\Omega_{f} = -E_\theta /
(\sqrt{\gamma}B^r)$, that approximates the value expected for a
split-monopole topology, $\Omega_{f} \simeq
\Omega_{h}/2$~\citep{Blandford1977, Komissarov2004b, Nathanail2014}.  At
the same time, the toroidal magnetic field respects the Znajek regularity
condition at the horizon, $B_\phi = \sqrt{\gamma} \left[ \beta^r D^\theta
  + \sin\theta \sqrt{\gamma} (\Omega_{f} - \Omega_{h}) B^r
  \right]$~\citep{Znajek77, Blandford1977, Uzdensky2005, Camilloni2022},
where $\beta^r$ is the shift vector and $\bm{D}$ is the electric field
measured by the FIDO~\citep{Komissarov2004b} (see the
\hyperref[sec:appendix]{Appendix} for details), and reverses polarity
across the equator.

\begin{figure*}
  \centering
  \includegraphics[width=1\textwidth]{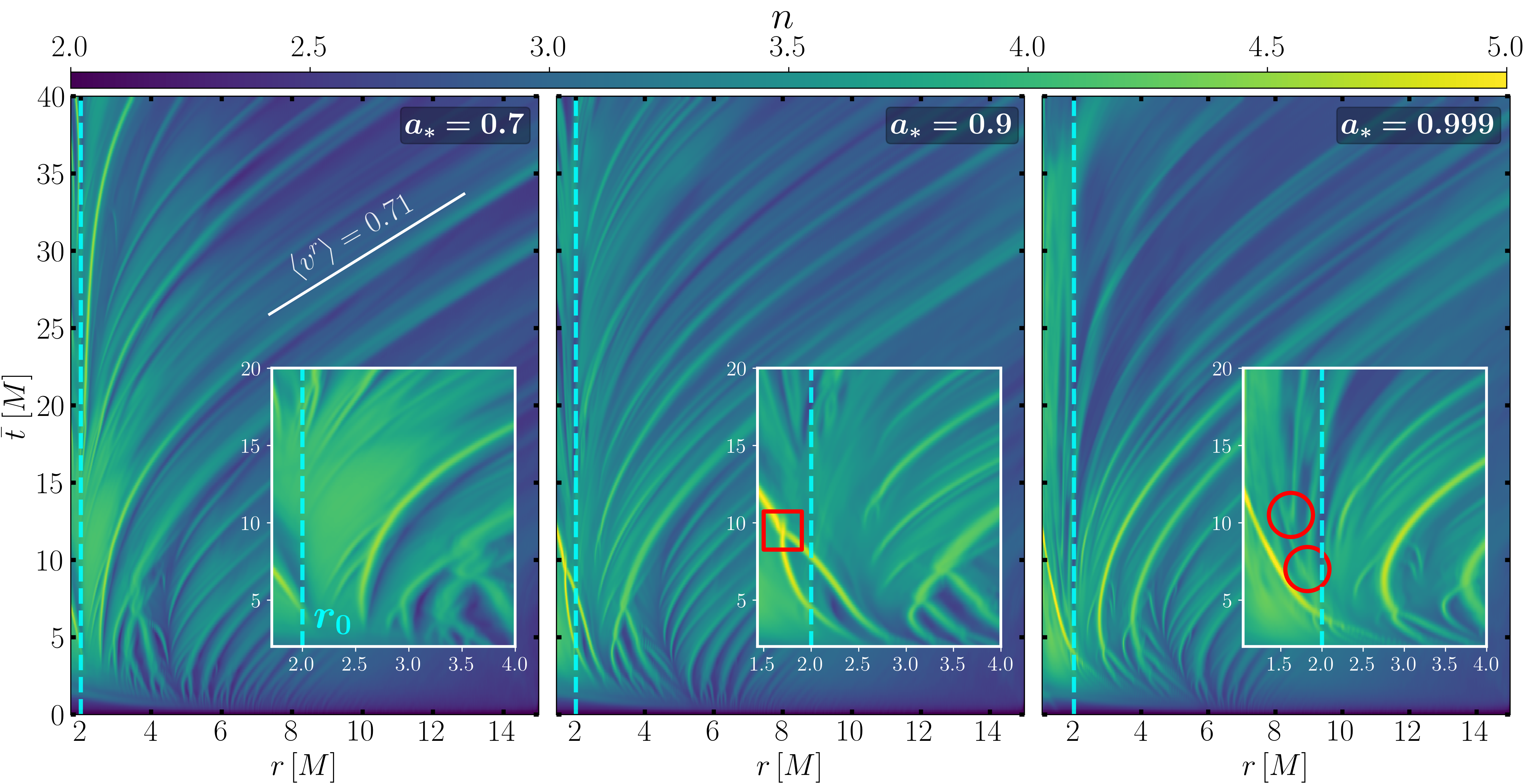}
  \caption{Spacetime diagrams of the total number density along the
    equatorial plane for BHs with spin $a_*=0.7$ (left panel), $a_*=0.9$
    (middle panel), and $a_*=0.999$ (right panel). The insets offer a
    magnification close to the event horizon and vertical dashed lines
    mark the location of the ergosphere $r_0$. Peaks of the number
    density track the motion of the plasmoids, which can collide (red
    square) and split (red circles) within the ergosphere; outgoing
    plasmoids are accelerated at relativistic speeds $v^r \sim 0.7$ (see
    slope).}
\label{fig:fig2}
\end{figure*}

Also shown in Fig.~\ref{fig:fig1} is the region inside the outer event
horizon $r_+$ (black-shaded area) and the position of the ergosphere, or
static limit, $r_0$ (cyan dashed line). We recall that under axisymmetry
and force-free quasi-stationarity, surfaces of superluminal rotation
appear for the magnetic-field lines and corresponding to the two zeroes
of the function $\mathcal{L}_{\rm ls}(\Omega_{f}, r, \theta) :=
g_{\phi\phi} \Omega_{f}^2 + 2 g_{t \phi}\Omega_{f} + g_{t t}=
0$~\citep{Komissarov2004b}, where $g_{\mu \nu}$ is the background Kerr
metric. The two zeros of $\mathcal{L}_{\rm ls}(\Omega_{f}, r, \theta)$
mark the inner and outer light surfaces and are reported with blue solid
lines in Fig.~\ref{fig:fig1}, with the former one being always located
inside the ergoregion \citep{Komissarov2004b}. Within the inner
light surface, massive particles cannot corotate and acquire instead
large radial infalling velocities at all latitudes, reaching Lorentz
factors up to $\Gamma \sim 200$ and comparable to those observed for the
particles accelerated by the magnetic reconnection along the equatorial
plane (see below).

Another important surface reported in Fig.~\ref{fig:fig1} is the
so-called stagnation surface, that is, the surface with vanishing radial
component of the three-velocity as measured from an asymptotic observer,
$\langle v^r \rangle=0$, where the brackets $\langle ~ \rangle$ indicate
the average of all the particles in a cell (the other components at the
equatorial stagnation surface are $\langle v^{\theta} \rangle \sim 0.0,
\langle v^{\phi} \rangle \sim 0.3$). We indicate it with a red solid line
and note that it is open at very large distances, as expected with a
plasma having a velocity component along the magnetic-field lines. Note
also how its location on the equatorial plane is modified by the
turbulent plasma dynamics resulting from reconnection and oscillates
within the ergosphere. Occasionally, the stagnation surface moves very
close to the event horizon, thus exposing large portions of the
ergosphere where the plasma has $v^r>0$ and can escape to large
distances. Finally reported in the right panel of Fig.~\ref{fig:fig1} is
the distribution of the cell-averaged Lorentz factor as measured by the
FIDO. As shown by the related colormap, the Lorentz factor is largest in
the current sheet and associated with the plasmoids, that can reach
velocities with $\Gamma \sim 100$. The configuration shown in
  Fig.~\ref{fig:fig1} refers to a time when a number of small plasmoids
  have merged into large ones (see also right panel of
  Fig.~\ref{fig:fig2} and the discussion below). In this sense, it does
  refer to a specific configuration within the timescale of our
  simulations, but we expect that configurations of this type will occur
  frequently in a steady-state production of plasmoids.

To better illustrate and understand the dynamics of plasmoids, we present
in Fig.~\ref{fig:fig2}, three spacetime diagrams for three representative
high-spin BHs ($a_*=0.7, 0.9, 0.999$ from left to right) over the time
range $0 \leq \bar{t}/M \leq 40$. More specifically, we report the
evolution in time of the total plasma density $n$ along the current sheet
(we average over a few cells around $\theta= \pi/2$ to minimise noise)
and in the radial region $r_{h} \leq r \leq 15\,M$. Also shown with a
cyan dashed line is the equatorial location of the ergosphere, $r_0=2M$,
while the three insets report a magnification of the region close to the
BH.

The spacetime diagrams clearly show the genesis of the plasmoids near the
equatorial plane and their successive evolution either towards the
horizon (leftwards trajectories), or to large distances (rightwards
trajectories). Note also how small plasmoids can have binary mergers,
sometimes multiple times (illustrated with a red square in the inset
within the middle panel), and that these can happen either inside the
ergosphere, or outside. In general, plasmoids that are born inside the
ergoregion move towards the event horizon and fall into the BH with a
relatively small radial velocity. On the other hand, those born outside
are ejected from the magnetosphere and accelerated to relativistic
velocities. The use of a spacetime diagram allows us to trivially
estimate the asymptotic velocities in terms of the slopes of the
rightwards trajectories (see white line in leftmost panel of
Fig.~\ref{fig:fig2}), from which we deduce that $\langle v^r \rangle \sim
0.7 $. Remarkably, because this velocity appears to be
essentially the same for all of the plasmoids and independently of the BH
spin, it suggests that a steady wind of plasmoids with radial velocity
$v^r \simeq 0.7$ should accompany the electrodynamics of plasma near BHs.

Importantly, note how in the case of the very rapidly spinning BH with
$a_*=0.999$ (rightmost panel in Fig.~\ref{fig:fig2}), and probably as a
combination of tidal forces and extreme rotation, the spacetime diagram
highlights the splitting of large plasmoids, leading to smaller plasmoids
moving in opposite directions (see red circles in rightmost inset). While
one plasmoid marches towards the event horizon, the other one seems to
leave the ergosphere with a small radial velocity but a large azimuthal
one. These conditions, that are present only in the ergosphere, are
reminiscent of the conditions expected to lead to a Penrose process via
magnetic reconnection~\citep{Asenjo2015, Comisso2021,
  Camilloni2025}. While in our simulations this splitting has been
measured only twice, this is, to the best of our knowledge, the first
time that this process is shown to take place. Longer simulations are
likely going to show more of these plasmoid-splittings, while larger
particle numbers and higher spatial resolutions will help to better track
the evolution of split plasmoids.

\subsection{Rate of magnetic reconnection}

One of the most important quantities characterising the dynamics of the
force-free plasma we have simulated is given by the magnetic reconnection
rate $\mathcal{R}$, which can be taken as the rate at which plasmoids are
produced along the current sheet. We here estimate the reconnection rate
in terms of the FIDO ``drift velocity'' that determines how fast
  the plasma (and thus the magnetic flux) is transported into the
  reconnecting region, defined as $V^i := \sqrt{\gamma}\, \eta^{ijk} D_j
B_k / B^2$, where $\eta^{ijk}$ is the Levi-Civita symbol. Using these
quantities, we compute the reconnection rate as \citep{ElMellah2021,ElMellah2023}
\begin{equation}
  \label{eq:rr}
  \mathcal{R} := \frac{1}{2}\Delta V^\theta\,,
\end{equation}
where $\Delta V^\theta$ represents the jump in the polar drift velocity
across the current sheet at the X-points, and $V^\theta$ shows a
  local maximum and minimum when crossing the current sheet, in the range
  $\theta = \pi/2 \pm \pi/20$. Also, we define the X-points as the
regions where the radial profile of $B^\theta$ changes sign and which is
located between two local maxima of the total number density taken to
represent the plasmoids. We should remark that alternative measures are
possible of the reconnection rate involving linear ratios of the electric
and magnetic fields (see, \eg \cite{Zenitani2001, Kagan2015, Liu2015,
  Crinquand2020b, Imbrogno2025}) and some authors also employ variants of
Eq.~\eqref{eq:rr} where FIDO and non-FIDO EM components are
used~\citep{ElMellah2021, Bransgrove2021}. Overall, the different
definitions are related and yield comparable estimates. Because
$\mathcal{R}$ exhibits a small but stochastic variation in time, we
compute the averaged reconnection rate $\langle \mathcal{R} \rangle$ by
sampling $\mathcal{R}$ with a frequency in time of $1 \,M$ at all the
X-points within a spatial bin of length $0.5 \,M$ on the equatorial
plane, and then averaging over the total number of reconnection points
contained in the bin.

\begin{figure*}
  \centering
  \includegraphics[width=0.75\textwidth]{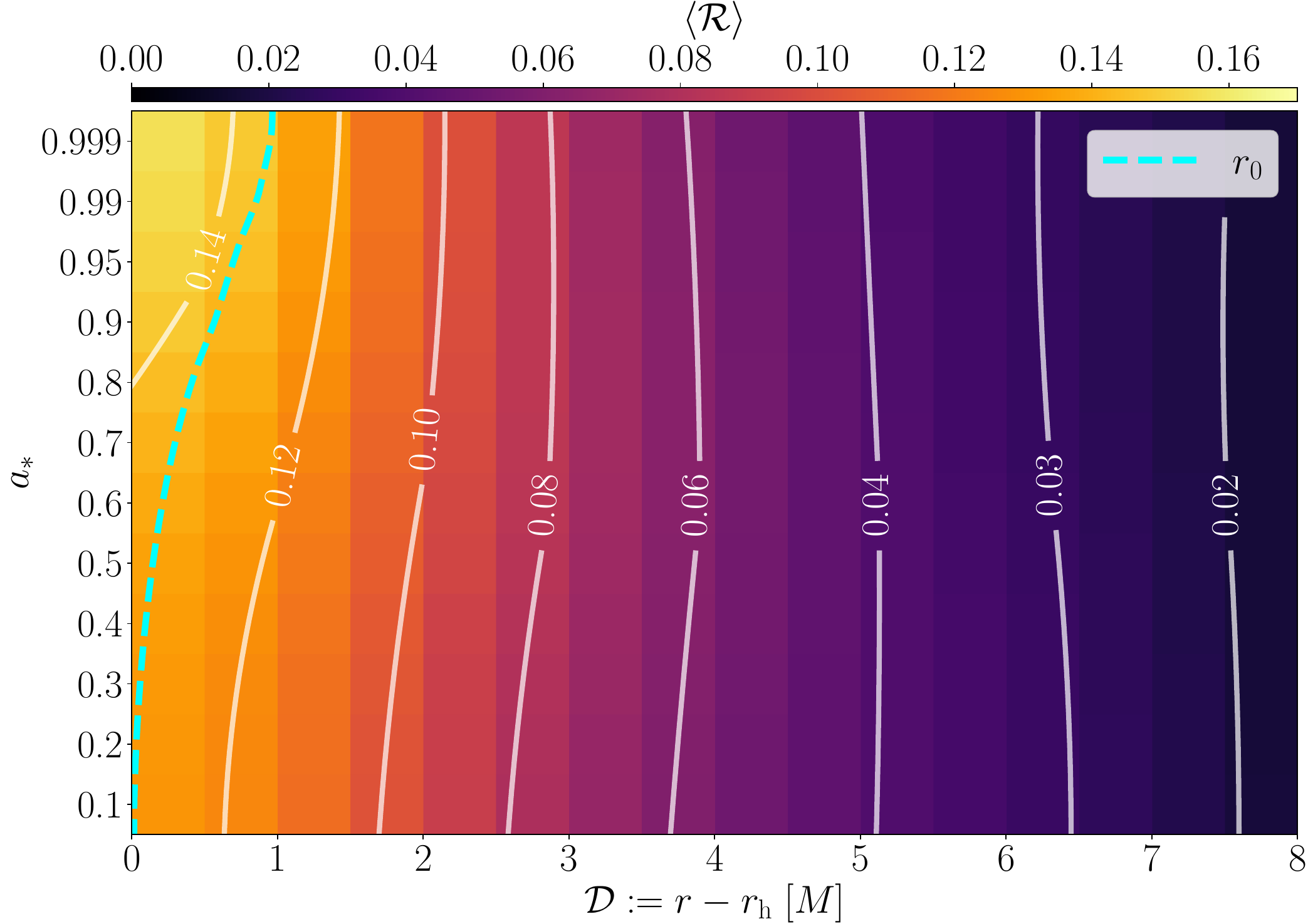}
  \caption{Averaged reconnection rate $\mathcal{R}$ reported as a
    function of the spin $a_*$ and the distance from the horizon
    $\mathcal{D}$. The average is computed in the time window $0 \leq
    \bar{t}/M \leq 40$ and with a spatial resolution of $0.5\,M$. The
    cyan dashed line shows the position of the ergosphere $r_0$ for
    different BH spins.}
\label{fig:fig3}
\end{figure*}

Since the plasma dynamics is strongly influenced by the background
spacetime, it is natural to expect that $\langle \mathcal{R} \rangle =
\langle \mathcal{R} \rangle (a_*,r)$~\citep{Asenjo2017}. As a result, in
Fig.~\ref{fig:fig3} we report the averaged reconnection rate $\langle
\mathcal{R} \rangle$ as a function of spin and distance on the equatorial
plane from the event horizon, $\mathcal{D} := r-r_{h}$. As expected from
Figs.~\ref{fig:fig1} and \ref{fig:fig2}, the majority of reconnection
events takes place near the BH and increases as a function of the
spin. For any spin, the largest rate is always attained in the ergosphere
(marked with a cyan dashed line) and reaches a maximum value of $\langle
\mathcal{R} \rangle \sim 0.15$ for the maximally spinning BH considered
with $a_* = 0.999$. Our results are in good agreement with other
measurements of the reconnection rate in collisionless plasmas near
BHs~\citep{ElMellah2021,Bransgrove2021,Crinquand2022}. At a distance
$\mathcal{D} \sim 7\,M$ the reconnection rate has decreased to $\langle
\mathcal{R} \rangle \simeq 0.02-0.03$, and becomes negligible at larger
distances.

Given the regular behaviour of $\langle \mathcal{R} \rangle(a_*,r)$, it
is possible to find a good analytic fit with the Ansatz
\begin{equation}
  \mathcal{R}_{\rm fit} = \frac{b_1(1+b_2 M^2\Omega_h^2)}{1+b_3
    {r^2}/{M^2}}\,,
\end{equation}
where the use of the angular frequency of the event-horizon $2M\Omega_h =
a_*/(1+\sqrt{1-a^2_*})$ captures better than $a_*$ the behaviour of the
data. With fitting coefficients $b_1= 0.183$, $b_2=-0.361$, and $b_3=
0.077$, we obtain a maximum relative error of $\lesssim 10\%$ for most of
the fitting domain in Fig.~\ref{fig:fig3}.

\subsection{Blandford-Znajek power}

\begin{figure*}
  \centering
  \includegraphics[width=0.8\textwidth]{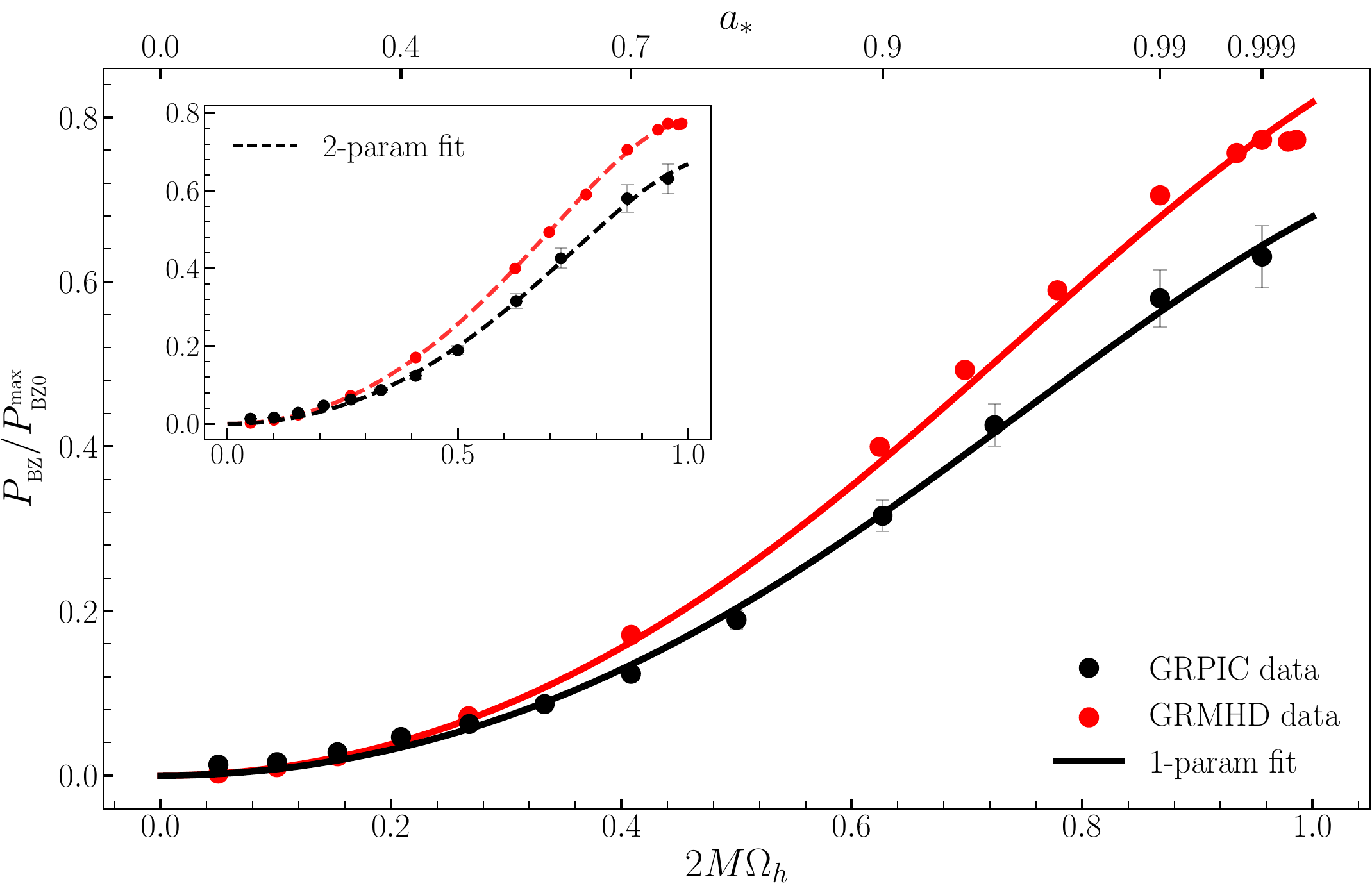}
  \caption{Blandford-Znajek luminosity $P_{_{\rm BZ}}$ normalised to the
    maximum value of the lowest-order (\ie quadratic) expression
    $P^{_{\rm max}}_{_{\rm BZ0}}$, as a function of the BH angular
    velocity (see top horizontal axis for a mapping in terms of the BH
    dimensionless spin) for all of our GRPIC simulations (black filled
    circles) and with the associated numerical errors. The black solid
    line shows the analytic expression~\eqref{eq:f_analytic} after
    rescaling the factor $\kappa$ to account for the specific
    magnetic-field topology (one-parameter fit). Also reported are the
    data from GRMHD simulations~\citep{Tchekhovskoy2010} (red filled
    circles) and the corresponding analytic expression suitably rescaled
    for $\kappa$ (red solid line). Shown in the inset with dashed lines
    of corresponding colours are the fits to the data in terms of the
    coefficients $\kappa$ and $\tilde{\beta}$ (two-parameter fit).}
\label{fig:BZP}
\end{figure*}

As anticipated, the BZ mechanism is an efficient process of conversion of
the rotational energy into EM energy taking place when a force-free
magnetosphere develops in BHs. As the preferred candidate mechanism to
explain the power of relativistic jets in AGNs and GRBs, it plays a
fundamental role in relativistic astrophysics. In what follows, we
present the first measurements of the BZ luminosity from a fully kinetic
GRPIC framework over the full range of the spin parameter. More
specifically, we calculate the power emitted via the BZ mechanism in our
split-monopole configuration by computing the Poynting flux over a
two-sphere close to the BH horizon
\begin{equation}
  P_{_{\rm BZ}}=\int \left(T^r_{~~t}\right)_{_{\rm
      EM}}\sqrt{-g}\,d\theta\, d\phi = 2\pi \int_0^\pi
  S^r\sqrt{\gamma}\,d\theta\,,
\end{equation}
with $\bm{T}_{_{\rm EM}}$ being the EM part of the energy-momentum
tensor, $S^r := \left( E_\theta H_\phi - E_\phi H_\theta \right) / (4 \pi
\sqrt{\gamma})$ is the radial component of the Poynting vector, and
$\bm{H}$ the magnetic field measured by an asymptotic observer (see
the \hyperref[sec:appendix]{Appendix} for details). The (time-averaged) results
of the GRPIC simulations are shown with black filled circles in
Fig.~\ref{fig:BZP}, with the corresponding error-bars.

Already in the original work by Blandford and Znajek, an analytic
expression was proposed to capture the dependence of the BZ power on the
BH spin. Such a dependence can be expressed generically as
\begin{equation}
  \label{eq:P_gen}
  P_{_{\rm BZ}}=\frac{\kappa}{4\pi}\,\Phi_{h}^2\, F(\Omega_{h})\,,
\end{equation}
where $\Phi_{h}$ is the (half-hemisphere) magnetic flux and $\kappa$ is a
scaling factor independent of the BH spin and related to the topology of
the magnetosphere; $\kappa_{sm}=1/6\pi\approx 0.053$ for an ideal
split-monopole magnetic field. The function $F(\Omega_{h})$ in
\eqref{eq:P_gen} expresses the spin-related contributions and for
slowly-spinning BHs $F(\Omega_{h}) = \Omega^2_{h} +
\mathcal{O}(M^4\Omega_h^4)$, so that the luminosity is given by the
leading-order quadratic term originally derived by Blandford and Znajek,
\ie $P_{_{\rm BZ0}} := (\kappa_{sm}/4\pi) \, \Phi_{h}^2 \, \Omega^2_{h} =
(1/24\pi^2) \, \Phi_{h}^2 \, \Omega^2_{h}$~\citep{Blandford1977}, and
$P^{_{\rm max}}_{_{\rm BZ0}}\approx 10^{-3}(\Phi_h/M)^2$ represents the
maximum value of the original BZ estimate attained for $M\Omega_h
= 1/2$. Using analytic techniques that combine perturbation theory with a
matched-asymptotic expansion scheme~\citep{Armas2020, Camilloni2022}, it
was recently possible to obtain high-order corrections in $F(\Omega_h)$
up to $\mathcal{O}(\Omega^8_h)$ and expressed as
%
%

\begin{equation}
\label{eq:f_analytic}
\begin{split}
F_{\rm an}(\Omega_{h}) = \Omega^2_{h} \left[ 1 + \tilde{\alpha}
      (M\Omega_{h})^2 + \tilde{\beta} (M\Omega_{h})^4
      + \tilde{\gamma} |M\Omega_{h}|^5 \right. \\
      \left. + \left(\tilde{\delta} + \tilde{\epsilon}
      \log|M\Omega_{h}|\right) (M\Omega_{h})^6\right] 
 + \mathcal{O}(M^9\Omega_{h}^9)\, ,
\end{split}
\end{equation}

where the coefficients in the expansion are given by $\tilde{\alpha} =
8{(67 - 6\pi^2)} / {45}\simeq 1.38$~\citep{Tanabe2008}, $\tilde{\beta}
\simeq -11.25$, $\tilde{\gamma} \simeq 1.54$, $\tilde{\delta} \simeq
11.64$, and $\tilde{\epsilon}\simeq 0.17$~\citep{Camilloni2022}. 

Expression~\eqref{eq:P_gen}, with $\Phi_h$ computed as the time-averaged
value of $1/2\int|B^r|\sqrt{\gamma}d\theta d\phi$ at the horizon, and
with the dependence~\eqref{eq:f_analytic} is reported as a black solid
line in Fig.~\ref{fig:BZP}, where a one-parameter fit was made to fix the
scaling factor $\kappa=0.041$ since in our simulations the turbulent
current sheet (slightly) distorts the magnetic-field topology away from
an ideal split-monopole. Note how the analytic expression provides an
excellent description of the GRPIC data at essentially all spins. To the
best of our knowledge, this is the first time that the BZ power has been
computed from kinetic simulations over the full range in spin and shown
to be in very good agreement with the analytic prediction (see
also~\cite{ElMellah2021} for a different magnetic-field configuration,
with spin limited to $a_* \geq 0.6$, and $P_{_{\rm BZ}}$ normalised to
the initial magnetic flux). Importantly, the analytic expression fits
very well also the data from two-dimensional GRMHD
simulations~\citep{Tchekhovskoy2010} (red filled circles in
Fig.~\ref{fig:BZP}), whose $P_{_{\rm BZ}}$ values are systematically
larger than the corresponding GRPIC ones (a result also found
in~\cite{Bransgrove2021}). For this data too, we have performed a
one-parameter fit for the scaling factor $\kappa=0.049$ (red solid line
in Fig.~\ref{fig:BZP}). For both GRPIC and GRMHD data, the variance
between the fit and the data is very small and is slightly better for the
former dataset. \cite{Tchekhovskoy2010} modelled the BZ luminosity
  truncating expression \eqref{eq:f_analytic} at
  $\mathcal{O}(\Omega^6_h)$, where the quadratic coefficient
  $\tilde{\alpha}$ was the only analytic one available at that
  time~\citep{Tanabe2008}, whereas the quartic coefficient
  $\tilde{\beta}$ was computed via a numerical fit. Similarly, we show
in the inset with corresponding colours and dashed lines the fits to the
data when performing two-parameter fits for the coefficients $\kappa$ and
$\tilde{\beta}$, and thus truncating Eq.~\eqref{eq:f_analytic} at
$\mathcal{O}(\Omega^6_h)$. Also in this case, the match with the data is
extremely good and similar to that presented in
Ref.~\cite{Tchekhovskoy2010}. Finally, we note that while we report
error-bars for the GRPIC data of $\sim 6\%$, this is not done for the
GRMHD data, as the errors are not available. However, we expect the data
from GRMHD simulations to have similar if not larger errors~(see, \eg
\cite{Porth2019_etal, Bransgrove2021}). Other tests of the BZ
  scaling are found in the literature, based on GRMHD simulations of BH
  accretion \citep{Tchekhovskoy2012, Narayan2022}, including
  two-temperature physics, heating, and radiative cooling
  \citep{Chael2025}.

\begin{figure*}
  \centering
  \includegraphics[width=0.85\textwidth]{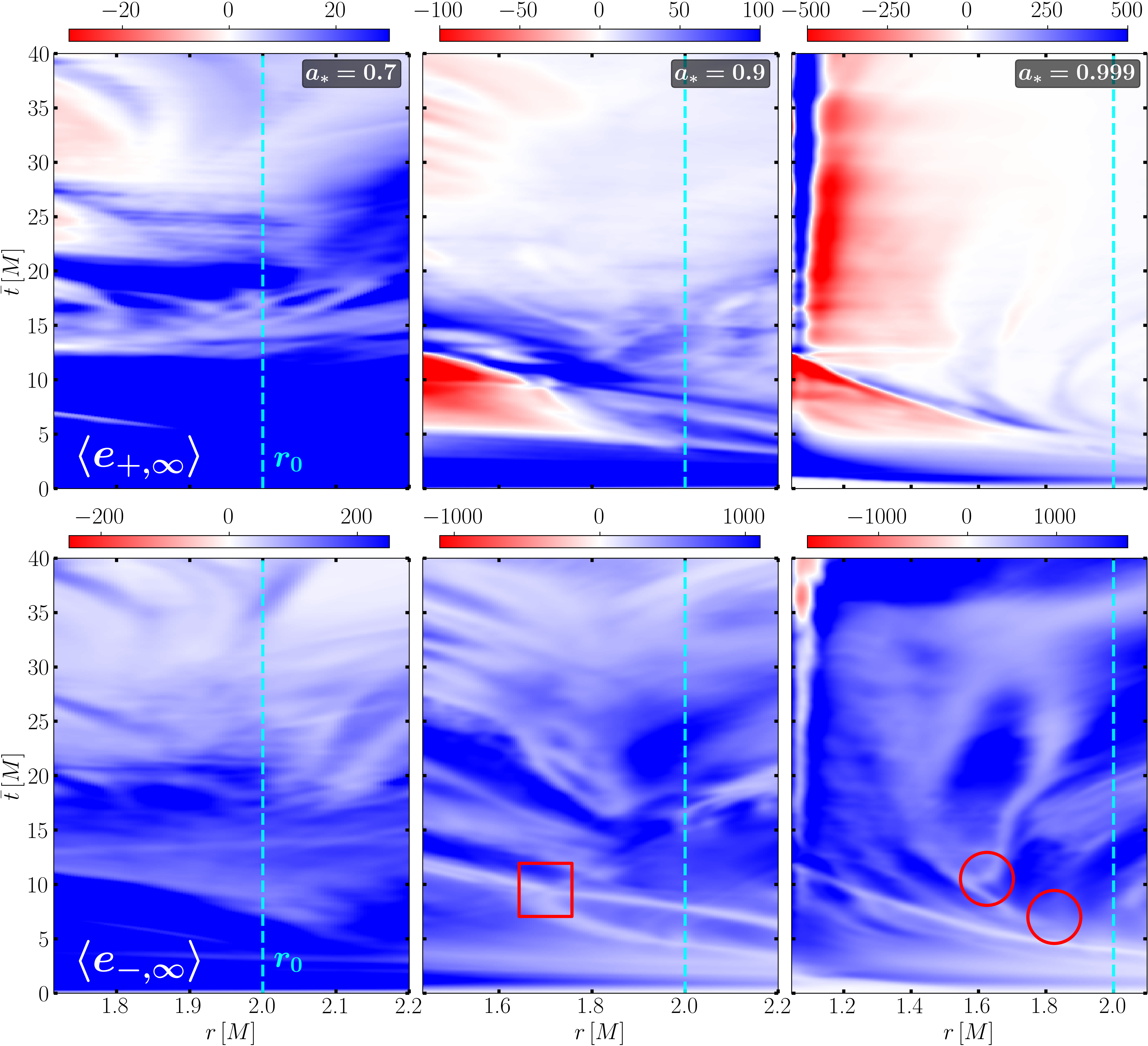}
  \caption{Spacetime diagrams for the average energy at infinity of
    positrons $\langle e_{+, \infty} \rangle$ (top row) and electrons
    $\langle e_{-, \infty} \rangle$ (bottom row) for the same simulations
    reported in Fig.~\ref{fig:fig2}. Note how negative-energy positrons
    are present even at moderate spins and their abundance and energies
    increases with the BH spin. In the case of electrons, only the
    highest spin leads to the generation of negative-energy electrons. As
    in Fig.~\ref{fig:fig2}, squares and circles are used to highlight the
    collision or splitting of plasmoids within the ergosphere.}
\label{fig_einf}
\end{figure*}

\subsection{Conditions for the Penrose process}

We conclude the presentation of our results with a discussion on the
particle energetics. Simulations with rapidly rotating BHs have revealed
the presence of particles with negative energy at infinity, \ie
$e_{\infty} := -u_t < 0$, where $u^\mu$ is the particle's four-velocity
(see, \eg ~\cite{Penrose69, Penna2015, Parfrey2019, ElMellah2021,
  Hirotani2021}). Under these conditions, a Penrose process is possible
whereby energy can be extracted from a Kerr BH when the negative-energy
particles cross the event horizon~\citep{Penrose69}. In
Fig.~\ref{fig_einf} we report spacetimes diagrams analogue to those shown
in Fig.~\ref{fig:fig2}, but where we concentrate on the (averaged) energy
at infinity of positrons, $\langle e_{+, \infty} \rangle$ (top row) and
of electrons, $\langle e_{-, \infty} \rangle < 0$ (bottom row; note that
the ranges differ from panel to panel). When focusing on the region close
to the horizon and inside the ergosphere, that is, the only region where
particles can show negative energies (red shading), it is simple to note
that large spins are needed to produce particles with large negative
energies and, indeed, the BH with $a_*=0.7$ (left column) show particles
with only moderate negative energies, while for the electrons $\langle
e_{-, \infty} \rangle \gtrsim 0$ (blue shading).

On the other hand, for the extreme-spin BH with $a_*=0.999$ (right
column), the ergosphere is wider and a large population of positrons
along the current sheet have negative energies. For $5 \lesssim \bar{t}/M
\lesssim 13$ we observe a persistent region of negative-energy positrons
near the horizon and entering it, thus suggesting that the Penrose
process is active when a large plasmoid enters the BH (see rightmost
panel of Fig.~\ref{fig:fig1} around $\bar{t}/M \sim 10$). Inflow of
negative-energy positrons can be seen very close to the event-horizon
also later on, but this is not steady and restricted to a thin strip near
the horizon. Note that at later times, \ie $\bar{t}/M \gtrsim 35$,
negative-energy electrons are produced near the event horizon and cross
it, thus suggesting that also electrons represent a channel to extract
energy from the BH. For less-extreme spins (middle column) this behaviour
is less pronounced, but negative-energy positrons are still present at
different times, with smaller negative energies matching the smaller BH
rotational energy, while no negative-energy electrons appear over the
timescale of our simulations. Indeed, a preliminary analysis, whose
  details will be presented in a companion
  paper~\citep{Camilloni2025b}, indicates that the outogoing power
  generated via the Penrose process by negative-energy positrons is about
  $10\%$ of that produced via the BZ mechanism. Overall, these results
-- including the distinct behaviour of electrons and positrons -- are in
good agreement with those presented in Ref.~\cite{Parfrey2019}, where
lower asymptotic energies were reported for the positrons and electrons.

Interestingly, when associating the trajectories of plasmoids with those
of the concentrations of the number density (as done in
Fig.~\ref{fig:fig2}), such trajectories actually refer mostly to
electrons with zero energies at infinity and are surrounded by regions
with $\langle e_{+, \infty} \rangle < 0$ or $\langle e_{+, \infty}
\rangle > 0$. This is particularly clear in the right panel of the top
row and suggests that in their propagation plasmoids ``leak'' positrons
of either negative or positive energy at infinity. This is the case also
for the splitting-plasmoids discussed in Fig.~\ref{fig:fig2} and whose
locations we report also in the right panel of the bottom row with red
circles. Also in this case, in fact, the plasmoid trajectory is
accompanied by regions of positrons with positive and negative
energies. While this behaviour is intriguing, simulations with
higher-resolution in particle number at extremal spins, that are
presently prohibitive, may be needed to reveal the complex relation
between plasmoid motion and the production of negative-energy particles.

\section{Discussion}
\label{concl}

Exploiting a newly developed general-relativistic kinetic code, we have
carried out the most comprehensive campaign of first-principle GRPIC
simulations of a classical scenario in relativistic astrophysics: a
rotating BH threaded by a split-monopole magnetic field. This
configuration is well-known to lead to the generation of a current sheet
near the equatorial plane and to magnetic reconnection, responsible for
the production of chains of relativistic plasmoids. Because they offer a
global view on the electrodynamics of BHs ranging from slowly rotating
ones over to BH near the extremal limit, our simulations have allowed us
to obtain a number of novel results.

First, we have been able to follow both in space and time the dynamics of
the plasmoids generated in the current sheet and to illustrate how they
either move towards the BH and are accreted or escape at large distances
at radial velocities $v^r \simeq 0.7$ for all BH spins. The study of
spacetime diagrams has also highlighted that plasmoids can collide and
generate larger ones, but also split if present in the ergosphere, thus
resembling the conditions expected to lead to a Penrose process via
magnetic reconnection. Second, using the large set of simulations, we
have been able to compute the reconnection rate under different
conditions and express it in terms of the BH spin and the distance from
the event horizon. The behaviour found for the reconnection rate is
regular and can be represented by an analytic function that can be used
in analytic models of magnetic reconnection in BH spacetimes.  Third, we
have measured the power associated with the Blandford-Znajek mechanism
and found excellent agreement with the high-order predictions coming from
perturbation theory. The match, which requires the calculation of a
single coefficient independent of the BH spin and related to the specific
magnetic-field topology, works equally well with data from GRMHD
simulations. To the best of our knowledge, this is the first time that
both GRPIC and GRMHD simulations of the BZ mechanism are shown to be
compatible with the analytic predictions and can be rescaled via a single
coefficient. Finally, the use of the spacetime diagrams and the analysis
of the plasmoids near the current sheet has highlighted that
negative-energy particles can be created inside the ergosphere and be
accreted by the BH, thus potentially leading to a Penrose process. While
positrons are much more likely to yield to negative-energy trajectories,
which are activated already at moderately high spins $a_* \simeq 0.7$ and
are abundant for $a_* \simeq 0.999$, also electrons have been found to
have negative energies for the most extreme spins. These results suggest
therefore that the electrodynamics of a rotating BH with a split-monopole
field may be responsible for both a BZ mechanism, and a
particle-based Penrose process.

While the results presented here are exciting and offer motivation for a
number of follow-up studies, we should also remark on some of their
limitations. First, they are restricted to two spatial dimensions as this
is the only route to afford high-resolution and long-term simulations
over a large range of BH spins. Clearly, a much richer dynamics is to be
expected when this analysis will be extended to three dimensions. Second,
while the particle and spatial resolution employed here are large and
sufficient to reach consistency in the results, some finer details, such
as those associated with the splitting of plasmoids, would benefit from
even more accurate calculations to address, for instance, the properties
of the negative-energy and outgoing plasmoid. Finally, the physical
conditions near a BH are surely more complex than those represented here
in terms of a pure pair-plasma. In addition to electrons and positrons,
in fact, protons and ions will be present and a full multi-species
description is necessary to study their dynamics. We plan to address at
least some of these aspects with future work.

\section{Acknowledgements}

We thank Luca Comisso and Ileyk El Mellah for insightful comments on the
manuscript, and Alejandro Cruz-Osorio for useful discussions during the
development of the code. Support comes from the ERC Advanced Grant
``JETSET: Launching, propagation and emission of relativistic jets from
binary mergers and across mass scales'' (Grant
No. 884631). L.~R. acknowledges the Walter Greiner Gesellschaft zur
F\"orderung der physikalischen Grundlagenforschung e.V. through the Carl
W. Fueck Laureatus Chair. The simulations were performed on the local ITP
Supercomputing Clusters Iboga and Calea, on the Goethe-HLR supercomputer,
and on HPE Apollo HAWK at the High Performance Computing Center Stuttgart
(HLRS) under the grant BNSMIC.

\newpage
\appendix
\renewcommand{\thefigure}{A\arabic{figure}}
\setcounter{figure}{0}

\section{Numerical Setup}
\label{sec:appendix}

We have performed GRPIC simulations using the newly developed Frankfurt
code \texttt{FPIC}, which models the dynamics of charged particles in a
stationary and axisymmetric spacetime described by the Kerr metric with a
dimensionless spin parameter $0 \leq a_* \leq 1$. The simulations are
performed in 2.5 dimensions, \ie in conditions where the particles have
motions in three spatial dimensions and the EM fields have three spatial
components but are evolved in a two-dimensional domain in response to the
underlying axisymmetry of the background
spacetime~\citep{Meringolo2025b}. More specifically, when considering a
spherical coordinate system $(r,\theta,\phi)$ and the BH spin axis to be
along the $\theta=0$ coordinate line, we assume invariance in the
azimuthal direction, so that $\partial_\phi \psi = 0$ for all quantities
$\psi$ in the code.

To ensure regularity of the spacetime metric at the BH horizon, the Kerr
metric is expressed in terms of the spherical Kerr-Schild coordinates
$(t,r,\theta,\phi)$ adopting a $3+1$ decomposition. In this way, both the
particles and the EM fields are evolved with respect to the proper time
of FIDOs, whose worldlines are orthogonal to the constant-$t$
slices~\citep{Macdonald1982, Komissarov2004b}. The line element is thus
given by $ds^2 = -\alpha^2 dt^2 + \gamma_{ij} (dx^i + \beta^i dt)(dx^j +
\beta^j dt)$, with $\alpha$ and $\beta^i$ respectively being the lapse
function and the shift vector (see, \eg \cite{Rezzolla_book:2013} for
explicit expressions of $\alpha, \bm{\beta}$, and $\bm{\gamma}$).

The EM fields are evolved according to the relativistic Maxwell-Faraday
and Maxwell-Ampere equations~\citep{Komissarov04}
\begin{eqnarray}
\label{eq:maxwell_1}
  &&\partial_t \boldsymbol{B} = -\nabla \times (\alpha \boldsymbol{D}
  + \boldsymbol{\beta} \times \boldsymbol{B})\,, \\
  &&\partial_t \boldsymbol{D} = \nabla \times (\alpha \boldsymbol{B} -
  \boldsymbol{\beta} \times \boldsymbol{D}) - 4 \pi \boldsymbol{J}\,,
\label{eq:maxwell_2}
\end{eqnarray}
with the auxiliary current $\boldsymbol{J} := \alpha \boldsymbol{j} -
\rho\,\boldsymbol{\beta}$, and where $\rho$, $\boldsymbol{j}$,
$\boldsymbol{B}$ and $\boldsymbol{D}$ represent respectively the (total)
electric charge and current densities, as well as the magnetic and
electric fields measured by the FIDOs. The EM fields as measured by an
asymptotic observer can always be recovered from FIDO EM fields via the
transformations
\begin{eqnarray}
  \label{eq:E_eq}
  \boldsymbol{E}&=\alpha \boldsymbol{D} + \boldsymbol{\beta} \times
  \boldsymbol{B}\,, \\
  \boldsymbol{H}&=\alpha \boldsymbol{B} - \boldsymbol{\beta} \times
  \boldsymbol{D}\,.
  \label{eq:H_eq}
\end{eqnarray}

The divergence-free condition for the magnetic field is preserved by
means of the staggered Yee grid, which ensures that $\partial_j
(\sqrt{\gamma} B^j)/\sqrt{\gamma} \sim 0$ up to the machine round-off
error~\citep{Yee66}. Note that when solving
Eqs.~\eqref{eq:maxwell_1}--\eqref{eq:H_eq}, the EM field components are
interpolated onto the same grid position via a metric-weighted linear
interpolation. In addition, we enforce the Maxwell-Gauss law for the
electric field by applying a divergence-cleaning
approach~\citep{birdsall2005plasma, Crinquand2020b} in which we employ an
iterative Jacobi method with 500 iterations and update the corrected
electric field via Eq.~\eqref{eq:maxwell_1} every 25 timesteps. In this
way, we can ensure that $||\partial_j (\sqrt{\gamma} D^j)/\sqrt{\gamma} -
4 \pi \rho||_{\infty} /|| 4 \pi \rho||_{\infty} \lesssim 10^{-3}$.

The general-relativistic equations of motion for the position and
four-velocity of the charged particles are given respectively by
(for completeness we restore here the speed of light)
\begin{align}
\label{eq:push_1}
\frac{1}{c} \frac{d x^i}{dt} &= \frac{\alpha}{\Gamma} \gamma^{ij} u_j -
\beta^i\,, 
\end{align}
\begin{widetext}
\begin{align}
\label{eq:push_2}
\frac{1}{c} \frac{d u_i}{dt} &= -\Gamma \partial_i \alpha + u_j \partial_i \beta^j 
- \frac{\alpha}{2 \Gamma} u_l u_m \partial_i \gamma^{lm}
 + \frac{q \alpha}{m c^2} \left[ \gamma_{ij} D^j +
\frac{\sqrt{\gamma} \, \eta_{ijk} \gamma^{jl} u_l B^k}{\Gamma} \right]\,.
\end{align}
\end{widetext}
where $\Gamma := \sqrt{1+\gamma^{ij}u_i u_j}$ is the particles Lorentz
factor measured by a FIDO, $m$ is the mass of the particles, and $q=\pm
e$ the charge.

\begin{table*}
\renewcommand{\thetable}{A}
\label{tab:table1}
\begin{tabular}{lrrrrrrrrrrrr}
  \hline
  \hline
  Simulation     & $\texttt{a.1}$ & $\texttt{a.2}$  & $\texttt{a.3}$  &
  $\texttt{a.4}$ & $\texttt{a.5}$ & $\texttt{a.6}$  & $\texttt{a.7}$  &
  $\texttt{a.8}$ & $\texttt{a.9}$ & $\texttt{a.95}$ & $\texttt{a.99}$ &
  $\texttt{a.999}$ 
  \\ \hline 
  $a_*$ & 0.1 & 0.2 & 0.3 & 0.4 & 0.5 & 0.6
  & 0.7 & 0.8 & 0.9 & 0.95 & 0.99 & 0.999 
  \\ 
  $\Omega_{h} ~ [M^{-1}]$ & 0.03 &
  0.05 & 0.07 & 0.10 & 0.13 & 0.17 & 0.20 & 0.25 & 0.31 & 0.36 & 0.43 &
  0.47 
  \\ 
  $r_{h} ~ [M]$ & 1.99 & 1.98 & 1.95 & 1.92 & 1.87 & 1.80 & 1.71
  & 1.60 & 1.44 & 1.31 & 1.14 & 1.04 
  \\ 
  $n_{_{\rm GJ}} ~ [M^{-3}]$ & 2.0 & 4.0 &
  6.1 & 8.3 & 10.7 & 13.3 & 16.2 & 19.9 & 24.9 & 28.8 & 34.5 & 38.0
  \\ 
  $n_0/n_{_{\rm GJ}}$ & 5.00 & 2.50 & 1.64 & 1.20 & 0.93 & 0.75 & 
  0.61 & 0.50 & 0.40 & 0.35 & 0.29 & 0.26
  \\ 
  $N_{p}/10^6$ & 37 & 61 & 82 & 106 & 127 & 157 & 182
  & 211 & 256 & 268 & 289 & 312 
  \\ \hline
\end{tabular}
\caption{Table of GRPIC simulations. Starting from the top, we report:
  the name of the Run, the dimensionless spin parameter $a_*$, the
  angular velocity $\Omega_{h}$, the event horizon radius $r_{h}$,
  the Goldreich-Julian density $n_{_{\rm GJ}}$, the normalised injected
  number density $n_0/n_{_{\rm GJ}}$, and the total
  number of particles $N_{p}$ evolved at $t \sim 90 \, M$, expressed
  in millions.}
\end{table*}

\texttt{FPIC} implements three different schemes to solve
Eqs.~\eqref{eq:push_1}--\eqref{eq:push_2}: an explicit fourth-order
Runge-Kutta (RK) integrator, a classical Boris scheme~\citep{Boris1973},
and a second-order implicit Hamiltonian method~\citep{Bacchini2018}. The
latter is the most accurate one but also computationally, while the Boris
and the RK schemes shown very similar results in terms of accuracy at a
fraction of the computational cost. As a results, although the RK
integrator is not symplectic, \ie the area of a given region of the phase
space is not guaranteed to be preserved over long timescales, (as
required by Liouville's theorem~\citep{Rezzolla_book:2013}), it has
represented for us the optimal approach in terms of accuracy and
computational costs. Note also that, at least for the Kerr spacetimes
considered here, both the metric components in
Eqs.~\eqref{eq:push_1}--\eqref{eq:push_2} and their derivatives are
computed analytically at the position of the particle. After updating the
positions and velocities of the particles, the charge and current
densities computed from the sources are moved to the Yee grid via a
metric-weighted 2D linear interpolation.

Particles reaching the outer boundary or falling into the BH are removed
from the ensemble, while they are reflected when passing across the poles
for axisymmetry (\ie with $u_{\theta}$ changing sign at the poles). As a
result, new particles need to be introduced to compensate for these
losses and maintain high the particle number density. We perform the
injection of new particles every $\Delta t_{\rm inj} = 0.01\,M$ and only
for those cells for which the multiplicity condition $\mathcal{M}<10$ is
satisfied. The chosen values for the frequency of injection and for the
multiplicity are adequate to fill the magnetosphere with $n \gg n_{_{\rm
    GJ}}$ (see Fig.\ref{fig:fig1}) without impacting the speed of the
simulations. Overall, the number of particles evolved at every time-step
can reach $N_{p} \simeq 3 \times 10^8$ for the most extreme spin values
(see Table~\ref{tab:table1}). Finally, the time-step $\Delta t$ is set by
the Courant-Friedrichs-Lewy (CFL) condition and given, in two dimensions,
by $\Delta t = \mathcal{C}[\beta^r/ \Delta r + \alpha \sqrt{\gamma^{rr} /
    \Delta r^2 + \gamma^{\theta \theta}/\Delta \theta^2}]^{-1}$, where
$\mathcal{C}=0.5$ is the CFL coefficient.

For each simulation, whose main properties are summarised in
Table~\ref{tab:table1}, the axisymmetric computational domain starts from
within the event horizon, thus in a region causally disconnected from the
rest of the domain, and covers a radial extent $0.95 \, r_{h} \leq r \leq
20\, M$ and a polar one $ 1/200 \leq \theta/\pi \leq 199/200$ via $N_r
\times N_\theta = 4196 \times 1024$ cells. An additional outer layer with
absorbing boundary conditions for the EM fields is applied at $r_{{abs}}
:= 0.9\, r_{{max}} = 18\, M$. In this layer, resistive terms are added to
Maxwell equations to damp the EM fields and ensure that the reflection of
EM waves at the outer boundary does not affect the inner computational
domain~\citep{Cerutti2015}.

\newpage



\end{document}